\documentclass[a4paper,11pt]{article}
\usepackage{pos}
\usepackage{subfigure}

\title{Exploring nucleon spin by Drell-Yan process}
\author*[a]{Wen-Chen Chang}

\affiliation[a]{Institute of Physics, Academia Sinica, Taipei, 11529, Taiwan}
\emailAdd{changwc@phys.sinica.edu.tw}

\abstract{Contributions of parton's orbital motion to the nucleon spin
  are encoded in the transverse-momentum dependent distributions
  (TMDs) and generalized parton distributions (GPDs). Currently these
  distributions are accessed through the lepton-induced processes such
  as semi-inclusive deep-inelastic scattering (SIDIS) and deeply
  virtual Compton Scattering (DVCS). The (un)polarized Drell-Yan
  process with hadron beams provides a unique and complementary probe,
  enabling critical tests of the universality properties of TMDs and
  GPDs in the space-like and time-like approaches. This talk reviews
  recent results of TMDs obtained from the Drell-Yan measurements and
  discusses future prospects for accessing nucleon GPDs through
  measurements of the exclusive pion-induced Drell-Yan process at
  J-PARC.}

\FullConference{
The 26th international symposium on spin physics (SPIN2025)\\
21–26 September, 2025\\
Qingdao (Tsingtao), Shandong Province, China\\}

\begin{document}
\maketitle

\section{Drell-Yan Process}

Drell-Yan process was proposed by Sidney Drell and Tun-Mao Yan in
1970~\cite{Drell:1970wh} to describe the large invariant-mass spectrum
of dimuon events in p+A collisions. The leading-order diagram is an
electro-magnetic process of same-flavor quarks and antiquarks
annihilating into a time-like virtual photons. The process could be
generalized into an annihilation of quarks and antiquarks of different
flavors into the $W$ bosons through the weak interaction. Due to its
specific sensitivity to the antiquark, Drell-Yan process is known to
provide the strongest constraint of intermediate-$x$ $\bar{d}/\bar{u}$
ratios of proton's parton distribution functions
(PDFs)~\cite{SeaQuest:2021zxb, Chang:2014jba, Peng:2014hta,
  Cocuzza:2021cbi}. The production at large transverse momenta mainly
contributed by the perturbative $qG$ and $GG$ QCD
processes. Theoretical description of this process reaches the best
perturbative accuracy~\cite{Chen:2022cgv}. Therefore, Drell-Yan
process becomes the candle measurements in the hadron colliders and
provides strong constraints on the proton PDFs~\cite{Peng:2014hta,
  NNPDF:2021njg}.

The virtual photon in the Drell-Yan process is of time-like property,
i.e. with a positive invariant mass, in contrast to the space-like
virtual photon of a negative mass in the deep-inelastic
scattering. The success of global analyses of proton PDFs provides a
strong support of the factorization assumption as well as the
universality of PDFs~\cite{NNPDF:2021njg}. The study of partonic
structures of protons has been extended from one-dimensional PDFs to
multi-dimensional TMDs and GPDs through which the contributions of
parton orbital motion to the nucleon spin can be accessed. In this
proceedings, we will present how the Drell-Yan process can be utilized
in studying these advanced soft QCD objects~\cite{Peng:2014hta},
emphasizing the importance of checking universality of them determined
by the space-like and time-like approaches.

\section{TMDs: Boer-Mulders and Sivers Functions}

The Boer-Mulders (BM) functions describe the correlation between
transverse spin and transverse momentum of quarks and gluon. It could
be extracted from the the azimuthal $\cos 2\phi$ asymmetry from
unpolarized semi-inclusive deep inelastic scattering (SIDIS)
process~\cite{Barone:2009hw}. However, the BM and Cahn effects
strongly couple with each other so that it is difficult to disentangle
these two contributions from the SIDIS asymmetry measurements.

The general expression for the unpolarized Drell-Yan angular
distribution is~\cite{Lam:1978pu}
%\begin{widetext}
\begin{equation}
\frac {d\sigma} {d\Omega} \propto 1+\lambda \cos^2\theta +\mu \sin2\theta
\cos \phi + \frac {\nu}{2} \sin^2\theta \cos 2\phi,
\label{eq:DY_ang}
\end{equation}
\noindent where $\theta$ and $\phi$ are the polar and azimuthal decay
angle of the $l^+$ in the dilepton rest frame. In 1978, Lam and
Tung~\cite{Lam:1978pu} derived a relation of the parameters $\lambda$
and $\nu$: $1-\lambda-2\nu=0$, to be preserved up to the NLO pQCD
effect. Fixed-target Drell-Yan measurements with proton beams were
found to follow this relation~\cite{NuSea:2008ndg}, whereas those with
pion beams exhibited significant deviations~\cite{NA10:1987sqk,
  E615:1989bda}. In 1999, Boer postulated the violation of Lam-Tung
(L-T) relation could be due to the BM effect and showed that the $\nu$
parameter is proportional to the convolution of the quark and
antiquark BM functions of the projectile and target
hadrons~\cite{Boer:1999mm}. It makes possible the extraction of BM
functions from the angular parameter $\nu$ of Drell-Yan process.

%The asymmetry in the azimuthal distribution can be understood by
%noting that the Drell-Yan cross section depends on the transverse
%spins of the annihilating quark and antiquark. Consequently, a
%correlation between the quark's transverse spin and its transverse
%momentum -- encoded in the Boer-Mulders functions -- gives rise to a
%preferred direction in transverse momentum space.

\begin{figure}[htbp]
\centering
\includegraphics[width=0.6\columnwidth]{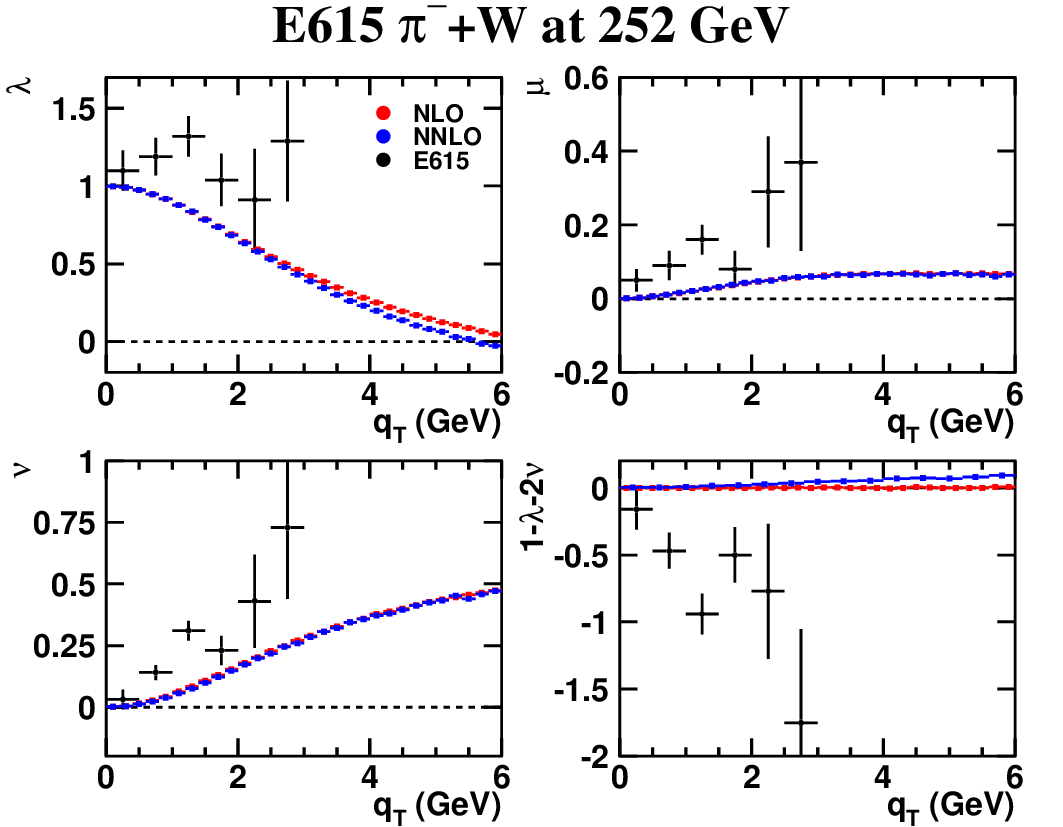}
\caption{Comparison of NLO (red points) and NNLO (blue points)
  fixed-order pQCD calculations with the E615 $\pi^-+W$ D-Y data at
  252 GeV~\cite{E615:1989bda} (black points) for $\lambda$, $\mu$,
  $\nu$ and $1-\lambda-2\nu$. Figure from~\cite{Chang:2018pvk}.}
\label{fig1_e615}
\end{figure}

In Ref.~\cite{Chang:2018pvk}, we contrasted the results of NLO and
NNLO fixed-order pQCD calculations of angular parameters with the NA10
and E615 measurements. Fig.~\ref{fig1_e615} shows the results of
$\lambda$, $\mu$, $\nu$, and the L-T violation, $1-\lambda-2\nu$, as a
function of transverse momentum $q_T$ from the fixed-order pQCD
calculations together with 252-GeV $\pi^- + W$ data from E615
experiment~\cite{E615:1989bda}. Overall, the calculated $\lambda$,
$\mu$ and $\nu$ exhibit distinct $q_T$ dependencies. At $q_T
\rightarrow 0$, $\lambda$, $\mu$ and $\nu$ approach the values
predicted by the collinear parton model: $\lambda = 1$ and $\mu = \nu
=0$. As $q_T$ increases, Fig.~\ref{fig1_e615} shows that $\lambda$
decreases toward its large-$q_T$ limit of $-1/3$ while $\nu$ increases
toward $2/3$. The $q_T$ dependence of $\mu$ is relatively mild
compared to $\lambda$ and $\nu$. Comparing the results of the NLO with
the NNLO calculations, $\lambda {\rm (NNLO)}$ is smaller than $\lambda
\rm{(NLO)}$ while $\mu$ and $\nu$ are very similar for NLO and NNLO.

The pQCD calculation predicts a sizable magnitude for $\nu$,
comparable to the data. Therefore, the pQCD contribution should be
taken into account in the future extraction of nonperturbative BM
functions from the Drell-Yan data of $\nu$~\cite{Lu:2009ip,
  Wang:2018naw}. The amount of L-T violation, $1-\lambda-2\nu$, is
zero in the NLO calculation, and turns to be nonzero and positive in
the NNLO calculation. The sign of L-T violation are opposite in the
NNLO pQCD predictions and the data and a further investigation is
desired. The most recent measurement of pion-induced Drell-Yan process
was performed by COMPASS experiment at
CERN~\cite{COMPASS:2010shj}. Preliminary results of angular
parameters~\cite{Lien:2022bwd} showed similar $q_T$ dependence,
deviation of $\nu$ parameter from pQCD calculation and violation of
L-T relation, consistent with previous measurements by NA10 and E615.

The Sivers functions is the most well-studied polarized TMDs. It
characterizes the correlation between the transverse momentum of
parton and the nucleon spin. It provides a naive picture of orbital
motion contributing to the nucleon spin. In the SIDIS process, a
non-zero Sivers asymmetries were clearly observed and a positive
(negative) Sivers function for $u$ ($d$) quark of the proton were
extracted from the global analysis~\cite{Anselmino:2012aa}. Because of
the opposite time directions of the Wilson lines in these processes,
perturbative QCD and TMD factorization predict a sign change of the
time-reversal-odd Sivers and BM functions between the SIDIS and
Drell-Yan processes~\cite{Collins:2002kn}. The validation of this
fundamental prediction is of essential importance for TMD physics.

Measurements of Sivers asymmetries in the polarized Drell-Yan
reactions were carried out by the COMPASS experiment in 2015 and
2018~\cite{COMPASS:2010shj}. The first measurement of Sivers
transverse beam asymmetries (TSA) of $A_{T}^{\sin \varphi_S}$ from the
2015 runs is $0.060 \pm 0.057 (\rm{stat}) \pm 0.040
(\rm{sys})$~\cite{COMPASS:2017jbv}, favoring the scenario of a sign
change. The following global analyses of Sivers functions with both
SIDIS and Drell-Yan data support the sign-change scheme. The combined
results of 2015 and 2018 runs for the beam asymmetries of $A_{T}^{\sin
  \varphi_S}$ is $0.070 \pm 0.037 (\rm{stat}) \pm 0.031 (\rm{sys})$
were published recently~\cite{COMPASS:2023vqt}. It is consistent with
the prediction incorporating the expected sign change with improved
statistic accuracy.

In addition, a transversity TSA $A_{T}^{\sin
  (2\varphi_{CS}-\varphi_{S})}$ asymmetries was found to be $-0.131
\pm 0.046 (\rm{stat}) \pm 0.047
(\rm{sys})$~\cite{COMPASS:2023vqt}. This asymmetries is proportional
to the convolution of BM of pion valence quarks and the transversity
of nucleon valence quarks, with an additional negative
sign~\cite{Bury:2020vhj}. Since the transversity of quarks in the
nucleon is of positive sign, the result suggests positive BM functions
for pion valence quarks. As mentioned before, the non-zero positive
$\nu$ parameter could be interpreted as the results of convolution of
BM functions of valences quarks of pion and
proton~\cite{Boer:1999mm}. We could derive that positive BM functions
for nucleon's valence quarks in the Drell-Yan process. Given the
negative BM functions of nucleon's valence quark in the SIDIS
process~\cite{Bury:2020vhj}, the negative transversity TSAs
$A_{T}^{\sin (2\varphi_{CS}-\varphi_{S})}$ support a sign change of
nucleon quark BM functions between the SIDIS and Drell-Yan
processes~\cite{COMPASS:2023vqt, Peng:2025etb}.

\section{GPDs: Exclusive Drell-Yan Process}

\begin{figure}[htbp]
\centering 
\includegraphics[width=0.9\columnwidth]{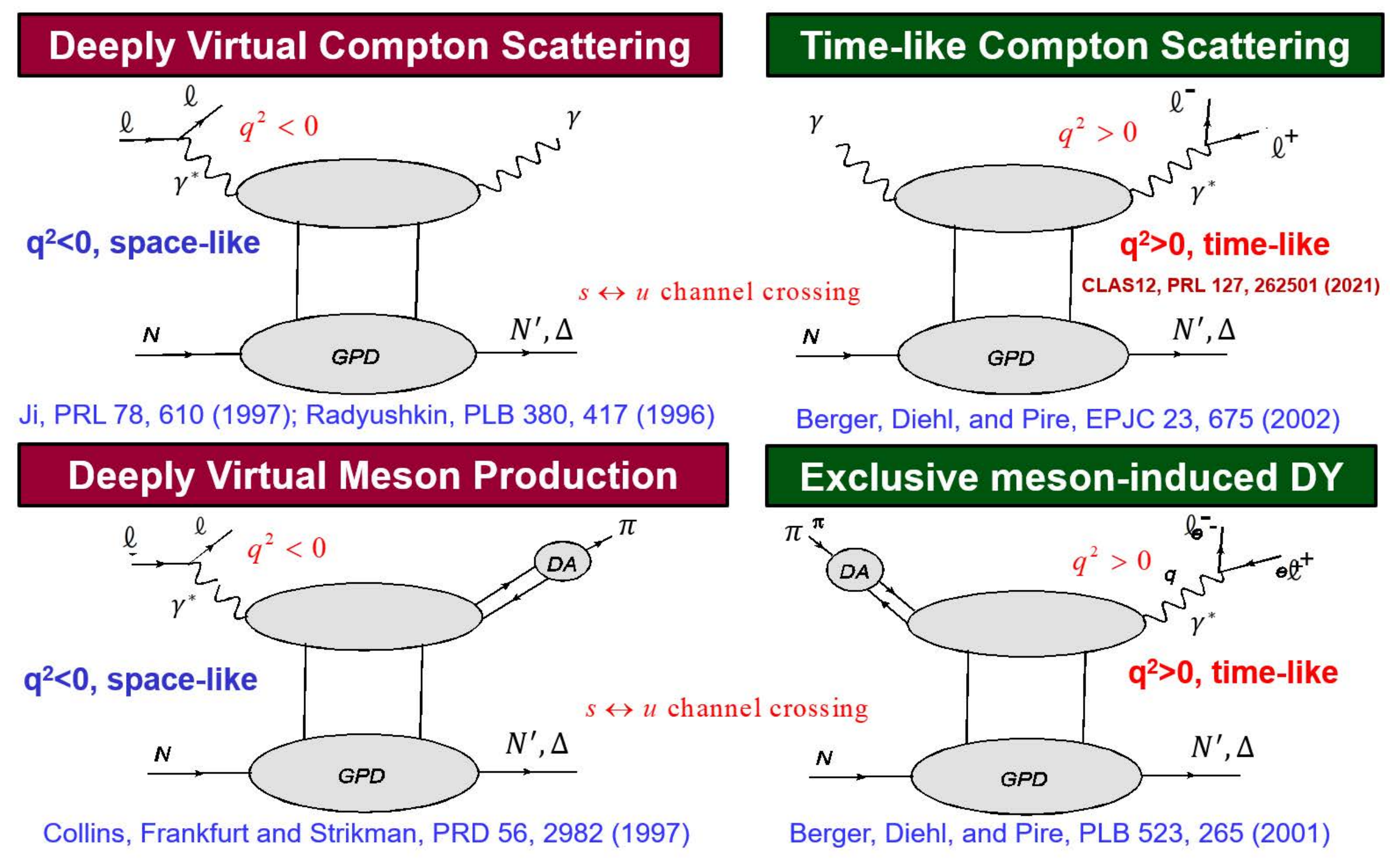}
\caption{Feynman diagrams illustrating DVCS, DVMP, TCS and exclusive
  Drell-Yan processes for accessing the nucleon GPDs.}
\label{fig2_GPD}
\end{figure}

Generalized parton distributions (GPDs) encode the correlation of
longitudinal momentum distribution and transverse size. As illustrated
in Fig.~\ref{fig2_GPD}, the GPDs have been accessed in deeply virtual
Compton scattering (DVCS) and deeply virtual meson production (DVMP)
processes with the lepton beams. In these approaches, the virtual
photons involved are of negative mass ($q^2 < 0 $), i.e. space-like
property. Through the $s$-$u$ channel crossing symmetry, GPDs could
also be extracted in the time-like processes of timelike Compton
scattering (TCS) and exclusive Drell-Yan process $\pi N \to \gamma^* N
\to l^+ l^- N$~\cite{Berger:2001zn,Goloskokov:2015zsa}. Recently the
TCS process was measured by CLAS experiment~\cite{CLAS:2021lky} and
the beam asymmetries are found consistent with the predictions based
on the GPDs extracted from DVCS and DVMP. These results provide strong
support for the universality of GPDs across these processes. In this
context, exclusive meson-induced Drell-Yan scattering offers a novel
and complementary approach to exploring GPDs using hadron beams.

In the limit of the large photon virtuality $Q^2 = -q^2 > 0$ at fixed
scaling variable, the amplitude of $\pi^-(q)+ {\rm p}(p) \to
\gamma^*(q')+ {\rm n}(p')$ can be written in terms of the
hard-scattering processes at parton level, combined with the DA
$\phi_\pi$ describing the formation of the pion from a $q\bar{q}$
pair, and also the nucleon GPDs, $\tilde{H}$ and $\tilde{E}$. The
leading-twist cross section is expressed as
follows~\cite{Berger:2001zn, Goloskokov:2015zsa}:
\begin{align}
\left.\frac{d\sigma_L}{dt dQ'^2}\right|_{\tau}
&= \frac{4\pi \alpha_{\rm em}^2}{27}\frac{\tau^2}{Q'^8} f_\pi^2\, \Bigl[ (1-\xi^2) |\tilde{\cal H}^{du}(\tilde{x},\xi,t)|^2 \nonumber \\
&- 2 \xi^2 \mbox{Re}\ \bigl( \tilde{\cal H}^{du}(\tilde{x},\xi,t)^* \tilde{\cal E}^{du}(\tilde{x},\xi,t) \bigr)
   -  \xi^2 \frac{t}{4 m_N^2}|\tilde{\cal E}^{du}(\tilde{x},\xi,t)|^2 \Bigr],
\label{eq_dcross}
\end{align}
where the fixed scaling variable is defined as $\tau = Q'^2 /(2, p
\cdot q)$, the skewness parameter as $\xi \simeq Q'^2 /(2s - Q'^2)$,
the Bjorken variable as $x = Q^2 /(2, p \cdot q)$, and the invariant
momentum transfer as $t = (p - p')^2$. The scaling variable $\tilde{x}
= - (q + q')^2 / \big[ 2 (p + p') !\cdot! (q + q') \big]$ can be
approximated as $\tilde{x} \simeq - Q'^2 /(2s - Q'^2) = -\xi$, and
$f_\pi$ denotes the pion decay constant. The subscript ``$L$'' of the
cross section indicates the contribution of the longitudinally
polarized virtual photon.

The convolution integral $\tilde{\cal H}^{du}$ involves two soft
objects: the GPD for $p\rightarrow n$ transition and the twist-two
pion distribution amplitude (DA) $\phi_{\pi}$. The expression of
$\tilde{\cal H}^{du}$ is given, at the leading order in $\alpha_s$,
by~\cite{Berger:2001zn}
\begin{align}
  \tilde{\cal H}^{du}(\tilde{x},\xi,t) &= \frac{8}{3} \alpha_s \int_{-1}^1
  dz\, \frac{\phi_\pi(z)}{1-z^2} \nonumber \\ 
  &\times \int_{-1}^1 dx
  \Bigl( \frac{e_d}{\tilde{x}-x- i\epsilon} - \frac{e_u}{\tilde{x}+x- i\epsilon}
  \Bigr) \bigl( \tilde{H}^{d}(x,\xi,t) - \tilde{H}^{u}(x,\xi,t)
  \bigr),
\label{eq_Hdu}
\end{align}
where $e_{u,d}$ are the electric charges of $u,d$ quarks in units of
the positron charge. The corresponding expression of $\tilde{\cal
  E}^{du}$ is given by (\ref{eq_Hdu}) with $\tilde{H}^q$ replaced by
the proton GPDs $\tilde{E}^q$. Due to the pseudoscalar nature of the
pion, the cross section (\ref{eq_Hdu}) receives the contributions of
$\tilde{H}$ and $\tilde{E}$ only, among the GPDs.

As shown in Eq.~(\ref{eq_dcross}), the term associated with
$|\tilde{\cal E}^{du}|^2 $ is multiplied by the momentum transfer
$|t|$ and thus the contribution of nucleon $\tilde{E}^{q}$ tends to be
suppressed at small $|t|$ compared to that of
$\tilde{H}^{q}$. Nevertheless, a remarkable feature of $\tilde{E}^{q}$
is that chiral symmetry ensures that $\tilde{E}^{q}(x,\xi,t)$ receives
a significant pion pole contribution for the ERBL region $|x| \le
\xi$, and, therefore, $\tilde{E}^{q}$ could play an important role at
small $|t|$ due to the proximity of the pion pole at $|t|=m_\pi^2$.

The leading-twist cross section (\ref{eq_dcross}) enters the four-fold
higher-twist differential cross sections for $\pi^- p \to \gamma^* n$
as~\cite{Goloskokov:2015zsa},
\begin{align}
\frac{d\sigma}{dt dQ'^2 d\cos\theta d\varphi} &= \frac{3}{8\pi} \bigl( \sin^2\theta \frac{d\sigma_L}{dt dQ'^2} + \frac{1+\cos^2\theta}{2} \frac{d\sigma_T}{dt dQ'^2} \nonumber \\ 
&+ \frac{\sin2\theta \cos\varphi}{\sqrt{2}} \frac{d\sigma_{LT}}{dt dQ'^2} + \sin^2\theta \cos2\varphi \frac{d\sigma_{TT}}{dt dQ'^2} \bigr),
\label{cadall}
\end{align}
with the angles $(\theta, \varphi)$ specifying the directions of the
decay leptons from $\gamma^*$. In Eq.~(\ref{cadall}), $d\sigma_T/(dt
dQ'^2)$ denotes the cross section contributed by the
transversely-polarized virtual photon. The $d\sigma_{LT}/(dtdQ'^2)$
and $d\sigma_{TT}/(dt dQ'^2)$ are the longitudinal-transverse
interference and transverse-transverse (between helicity $+1$ and
$-1$) interference contributions, respectively. The angular structures
of these four terms, characteristic of the associated virtual-photon
polarizations, allow us to separate the contribution of the
leading-twist cross section $d\sigma_{L}/ (dt dQ'^2)$ from the
measured angular distributions of dilepton pairs.

To carry out the measurements of exclusive Drell-Yan process, an
optimized beam momentum of pion beam is estimated to be around 10-20
GeV, based on several considerations. First, predictions for the
production cross sections at 10-20 GeV lie in the range of 1-100 pb
~\cite{Berger:2001zn,Goloskokov:2015zsa}, and the cross sections fall
rapidly decrease rapidly with increasing beam momentum. At beam
momenta below 10 GeV, the dimuon mass $Q'$ is insufficient to ensure
the validity of a hard scattering process. Second, a missing-mass
technique is required to guarantee exclusivity without detecting the
final-state neutron. Achieving good resolution for missing-mass
reconstruction precludes the conventional way of placing a hadron
absorber before the spectrometer, in the Drell-Yan measurement. At the
same time, to avoid large charged-track multiplicities in the
spectrometer, the beam momentum cannot be too high. Taken together,
these considerations make it feasible to perform the measurement using
the planned $\pi$20 high-momentum beamline in the hadron hall
extension project at J-PARC \cite{Aoki:2021cqa}.

In the $\pi$20 high-momentum beamline, 5-20 GeV high-flux secondary
pion and kaon beams with 0.1\% momentum resolution will be
delivered. Assuming the case of 30-kW primary protons on a 60-mm-long
platinum target (15-kW loss), the expected intensities of negative
pions, kaons, and antiprotons per spill (5.2-second spill cycle) are
in the order of $10^8$, $10^6$ and $10^5$ respectively. The first
approved experiment to utilize the secondary beams is E50/MARQ
experiment~\cite{Shirotori:2025MS}. Through a spectrometer of high
momentum resolution, the strange $\Omega$ and charmed baryon
spectroscopy are to be measured through the missing-mass technique for
investigating the diquark structure of heavy quark system. By
incorporating a muon-identification system, consisting of absorber
materials and tracking planes downstream of the spectrometer, the
exclusive Drell-Yan processes $\pi, K,N \to \gamma^* N \to \ell^+
\ell^- N, \Lambda$ can be studied through the detection of the dimuon
pair combined with cuts on the missing-mass spectra.

\begin{figure}[htbp]
\centering
\hspace{-0.5cm}
\subfigure[]
{\includegraphics[width=0.33\textwidth]{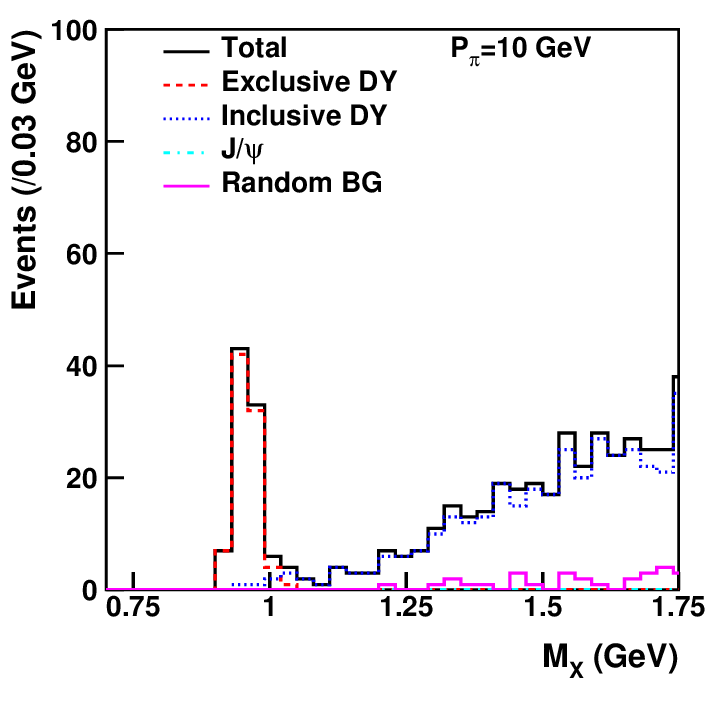}
\label{fig:mmass_1}}
\hspace{-0.5cm}
\subfigure[]
{\includegraphics[width=0.33\textwidth]{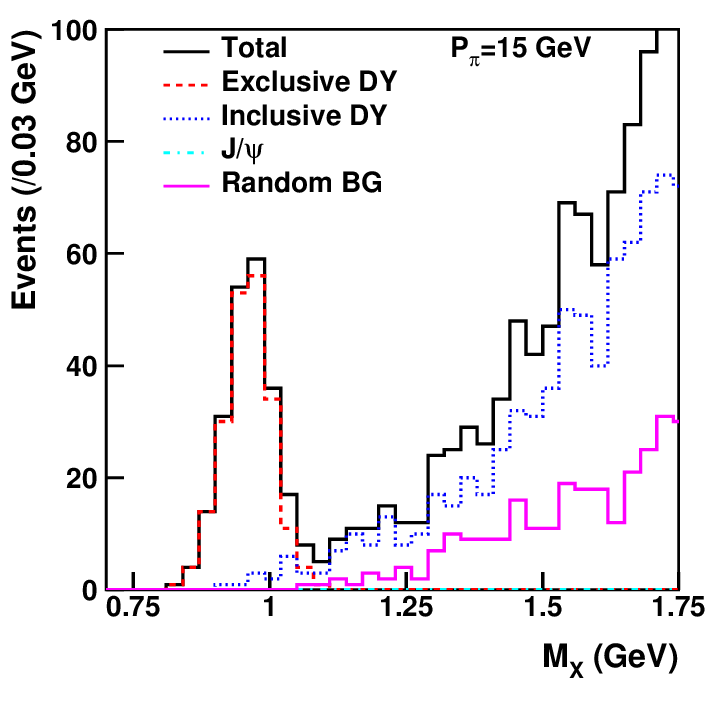}
\label{fig:mmass_2}}
\hspace{-0.5cm}
\subfigure[]
{\includegraphics[width=0.33\textwidth]{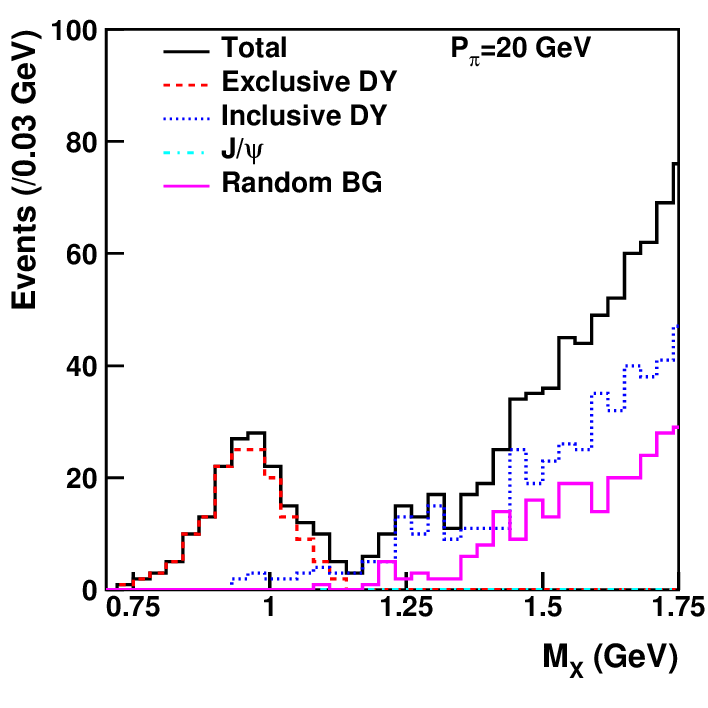}
\label{fig:mmass_3}}
\caption{Monte-Carlo simulated missing-mass $M_{X}$ spectra of the
  $\mu^{+}\mu^{-}$ events with $M_{\mu^{+}\mu^{-}} > 1.5$ GeV and
  $|t-t_{0}|<0.5$ GeV$^2$ for $P_{\pi}$=10, 15, and 20 GeV. Lines with
  different colors denote the contributions from various sources. The
  GK2013 GPDs is used for the evaluation of exclusive Drell-Yan
  process. Figures from~\cite{Sawada:2016mao}.}
\label{fig3_mmass}
\end{figure}

A feasibility study of measuring the exclusive pion-induced Drell-Yan
process $\pi^- p \to \gamma^* n \to \mu^+\mu^- n$ using E50 detector
configuration together with $\mu$ID system was done in
Ref.~\cite{Sawada:2016mao}. The estimated total cross sections for the
exclusive and inclusive Drell-Yan events for the dimuon mass
$M_{\mu^{+}\mu^{-}} > 1.5$ GeV and the $|t-t_{0}|<0.5$ GeV$^2$ are
about 10-20 pb and 2-3 nb, respectively. The $t_{0}$ denotes the
limiting value of 4-momentum transfer square $t$. In the range of beam
momentum 10-20 GeV, the total hadronic interaction cross sections of
$\pi^- p$ is about 20-30 mb while the production of $J/\psi$ is about
1-3 nb. Using GK2013 GPDs~\cite{Kroll:2012sm} for the exclusive
Drell-Yan process, the Monte-Carlo simulated missing-mass $M_{X}$
spectra of the dimuon events for $P_{\pi}$=10, 15, and 20 GeV is shown
in Fig.~\ref{fig3_mmass}. Lines with different colors denote the
contributions from various sources: exclusive Drell-Yan (red, dashed),
inclusive Drell-Yan (blue, dotted), $J/\psi$ (cyan, dash-dotted) and
random background (purple, solid), respectively. With 50-day run time
and 15-GeV pion beams, the statistic power is good enough to
differentiate the cross sections predicted by two GPD models. This
measurement is essential for verifying the universality of GPDs in
time-like processes, in close analogy to the comparison between TCS
and DVCS.

Due to the budget constraint, the $\pi$20 beamline is planned to be
realized in three phases: (1) Phase 1: flux of $10^5$/spill from the
interaction of primary 30-GeV proton beams with the Lamberton bending
magnet; (2) Phase 2: flux of a few $10^6$/spill by implementing a thin
production target and a swinger magnet; (3) Phase 3: flux of
$6x10^7$/spill by implementing a gas-tight production target and
completion of radiation shield. A pioneering measurement of the
secondary beam in Phase 1 was carried out by the T106 experiment in
January 2025. Encouraging results of identifying the secondary $\pi$
and $K$ were obtained and the flux was consistent with the
simulations~\cite{Noumi:2025yI}.

Another way of measuring nucleon transition GPDs through two-to-three
hard hadronic process with the proton beams has been proposed in
Ref.~\cite{Kumano:2009he}. To ensure a hard process, the transverse
momentum of rescattering baryons and mesons are required to be
large. This process can be categorized as one of the single
diffractive hard exclusive processes (SDHEPs) for the study of
GPDs~\cite{Qiu:2022pla} as listed in Table.~\ref{tab_SDHEPs}. The
measurement of $t'$ dependence of cross sections could be used to
explore the $x$-dependence of GPDs. A feasibility study is underway to
assess whether this measurement can be carried out within the ongoing
J-PARC E16 experiment or using a simplified mini-MARQ
spectrometer~\cite{Tomida:2025Y7}.

\begin{table}[hbtp]   %\footnotesize
%\setlength\tabcolsep{2pt}
%\addtolength{\tabcolsep}{2pt}
\centering
%\begin{center}
\begin{tabular}{|c|c|c|c|c|}
\hline
 $B$ & $h$ & $C$ & $D$ & Process \\
\hline
$\gamma^*$ & N & $\gamma$ & & DVCS \\
$\gamma^*$ & N & $\pi, \phi, J/\psi$ & & DVMP  \\
$\gamma^*$ & N & $l^+$ & $l^-$ & Double DVCS \\
$\gamma$ & N & $l^+$ & $l^-$ & TCS \\
$\gamma$ & N & $\gamma$ & $\pi, \phi, J/\psi$ & \\
$\gamma$ & N & $\gamma$ & $\gamma$ & \\
$\gamma$ & N & $\pi, \phi, J/\psi$ & $\pi, \phi, J/\psi$ & \\
\hline
\end{tabular}
\begin{tabular}{|c|c|c|c|c|}
\hline
 $B$ & $h$ & $C$ & $D$ & Process \\
\hline
$\pi$ & N & $l^+$ & $l^-$ & ~\cite{Berger:2001zn,Goloskokov:2015zsa} \\
$\pi$ & N & $\gamma$ & $\pi, \phi, J/\psi$ &  \\
$\pi$ & N & $\gamma$ & $\gamma$ &  \\
$\pi$ & N & $\pi, \phi, J/\psi$ & $\pi, \phi, J/\psi$ &  \\
N & N & $\pi$ & N & \cite{Kumano:2009he} \\
\hline
\end{tabular}
%\end{center}
\caption {Single diffractive hard exclusive processes (SDHEPs) $B + h
  \rightarrow C + D + h'$ proposed for the study of
  GPDs~\cite{Qiu:2022pla}.}
\label{tab_SDHEPs}
\end{table}

\section{Summary}

The Drell-Yan process has long been a powerful time-like probe of the
partonic structure of nucleons. Measurements in fixed-target
experiments via virtual-photon production, and in collider experiments
via $W$ and $Z$ boson production, have provided strong constraints on
the proton's PDFs. Its impact on nucleon spin physics, in terms of the
TMD Boer-Mulders and Sivers functions, can be accessed through
measurements of $\cos 2\phi$ angular distributions and transverse spin
asymmetries with transversely polarized targets. A crucial test of the
universality of Sivers functions between space-like and time-like
processes has already been achieved. Further studies of GPDs can be
pursued via the exclusive Drell-Yan process and other hard hadronic
reactions using pion or proton beams at J-PARC. Such measurements will
not only open new avenues for accessing GPDs but also provide a
fundamental test of their universality, analogous to what has been
established for PDFs and TMDs.

%\begin{thebibliography}{99}
%\bibitem{...}
%\end{thebibliography}

%JHEP
\bibliographystyle{JHEP} 
\bibliography{ref}

\end{document}